\documentclass[graphics,floatfix, footinbib,tightenlines,nobibnotes, aps, prb, twocolumn]{revtex4-1}
\usepackage{amsmath,amssymb,wasysym}
\usepackage{graphicx}
\usepackage{dcolumn}
\usepackage{bm}
\usepackage{braket}
\usepackage{subcaption}
\usepackage{verbatim}
\usepackage{float}
\usepackage{bbold}
\usepackage{color}
\usepackage{xcolor}
\usepackage{relsize}
\usepackage{amsthm}
\usepackage{enumerate}
\usepackage{soul,xcolor}
\usepackage[T1]{fontenc}
\usepackage{adjustbox}
\usepackage[colorlinks=true ,urlcolor=blue,urlbordercolor={0 1 1}]{hyperref}

\newcommand{\be}{\begin{equation}}
\newcommand{\ba}{\begin{align}}
\newcommand{\ee}{\end{equation}}
\newcommand{\bea}{\begin{eqnarray}}
\newcommand{\eea}{\end{eqnarray}}
\newcommand{\beq}{\begin{equation}}
\newcommand{\eeq}{\end{equation}}
\newcommand{\beqn}{\begin{eqnarray}}
\newcommand{\eeqn}{\end{eqnarray}}

\begin{document}
\widetext

\title{Ancilla wavefunctions of Mott insulator and pseudogap metal through quantum teleportation }
\author{Boran Zhou}
\author{Ya-Hui Zhang}
\affiliation{William H. Miller III Department of Physics and Astronomy, Johns Hopkins University, Baltimore, Maryland, 21218, USA}
\date{\today}

\begin{abstract}
Weak Mott regime with finite U is a wonderful region to search for quantum spin liquid, but it is challenging to write down a wavefunction capturing both spin liquid and charge fluctuations. Conventional methods using complicated Jastrow  factors have difficulties when the underlying spin liquid has a non-trivial projective symmetry group (PSG).  To cure this problem, here we provide a new  class wavefunction for Mott insulator through quantum teleportation using ancilla qubits.  We primarily focus on half filling of the fermionic  Hubbard model. We will prove that  a single variation parameter $\Phi$ in our wavefunction tunes the Mott charge gap continuously.  On a generic lattice, we show that the wavefunction  at $\Phi=+\infty$ recovers the familiar Gutzwiller projectived wavefunction at infinite U. The wavefunction at $\Phi=\frac{U}{2}$  is equivalent to applying the inverse Schrieffer–Wolff transformation at linear order of $t/U$, as expected in large but finite U regime.  From a gauge theory description we can show that the wavefunction has an electronic sector decoupled from a spinon sector describing localized spin moments.  The charge gap $\Delta_c$ can be shown to be $2\Phi$ and we conjecture that the wavefunction works well down to the regime with small charge gap on a generic lattice. We represent the wavefunction using tensor network and numerically confirm this conjecture in  one dimension.  Beyond the numerical power, the ancilla wavefunction also provides a new conceptual picture to understand the bandwidth tuned metal insulator transition. In this new framework, there can in principle exist a narrow region of fractional Fermi liquid (FL*) phase  between the usual Fermi liquid and the Mott insulator, a scenario which is not captured by the conventional slave rotor theory and thus was usually outlooked.    
\end{abstract}

\maketitle

\textbf{Introduction} Mott physics is at the center of condensed matter physics, due to its close relation to many important topics including high temperature superconductor\cite{lee2006doping} and quantum spin liquids\cite{savary2016quantum,norman2016colloquium,zhou2017quantum,broholm2020quantum}.  The ideal model to capture Mott physics is the fermionic Hubbard model, which is known to be relevant to a variety of condensed matter systems, ranging from conventional solid state material such as cuprates\cite{lee2006doping}, to moir\'e systems\cite{wu2018hubbard,tang2020simulation,regan2020mott,wang2020correlated,kennes2021moire} and to optical lattices in cold atom systems\cite{esslinger2010fermi}. Despite intense studies for decades, there are still many unsolved problems in this simple model.

At filling of one electron per site ($1/2$ filling per spin), a Mott insulator can be formed if the Hubbard U is much larger than the hopping $t$. When $U$ is infinite, projected Gutzwiller wavefunction is known to work well\cite{lee2006doping} and is a powerful tool to study quantum spin liquids\cite{anderson1987resonating,anderson2004physics,paramekanti2001projected,capriotti2001resonating,capello2005variational,ran2007projected,becca2017quantum}. However, there is no simple extension of the wavefunction down to finite U. One natural choice with a generalized projector $P_G=\prod_i (1-\alpha n_{i;\uparrow} n_{i;\downarrow})$ gives a metallic phase once $\alpha<1$.  To capture a Mott insulator with finite charge gap, one may need to add a doublon-holon binding factor\cite{guertler2014kagome} or more complicated  Jastrow factor $J_d=e^{-\frac{1}{2} V_{ij} n_i n_j}$\cite{capello2005variational,capello2006unconventional} even with backflow terms\cite{tocchio2011backflow}. One drawback of this approach is that the Mott gap is not controlled by a single parameter.   An even more serious issue arises when the underlying state is a spin liquid with a flux per unit cell (a projective translation symmetry\cite{wen2002quantum}) such as a Dirac spin liquid\cite{wen2002quantum} or certain chiral spin liquid \cite{song2021doping}. In this case the Jastrow factor wavefunction actually breaks the translation symmetry because the fermion here plays the role of both spinons and charge and the density fluctuations feel the non trivial PSG as real symmetry breaking pattern\footnote{There is no such issue in the Gutzwiller projected wavefunction at infinite U because the density fluctuation is completely frozen there.}.
However, a symmetry Dirac spin liquid or chiral spin liquid with finite charge gap obviously exist. For example, a translationally invariant chiral spin liquid (CSL) was observed from unbiased density matrix renormalization group (DMRG) calculation in the finite U regime\cite{szasz2020chiral}.  The  CSL here is identified as a U(1)$_2$ state and should have a $\pi$ flux ansatz in parton construction\cite{song2021doping}. However, in variational study using the standard Jastrow factor, this more natural ansatz  was ruled out\cite{tocchio2021hubbard} because it breaks the translation symmetry due to the artifact of the Jastrow factor approach. Apparently representing a spin liquid with non-trivial PSG in the finite U regime is a fundamental problem which does not have a good solution.

In this paper, we propose a new class of variational wavefunction which solves  the above issues.  In this wavefunction, the Mott charge gap is controlled by a single parameter $\Phi$ and we have explicit charge and spin separation.  The key is to introduce ancilla degrees of freedom and then project them out in the end. The ancilla wavefunction was introduced previously by one of us from a phenomenological perspective for the pseudogap metal in hole doped cuprates\cite{zhang2020pseudogap}. But its validity and energetics in a microscopic Hamiltonian has not been carefully benchmarked. In this paper we mainly focus at half filling and show analytical proofs that the wavefunction recovers the familiar Gutzwiller projected wavefunction at infinite $U$ and is equivalent to applying inverse inverse Schrieffer–Wolff transformation at linear order of $t/U$.  We also represent the wavefunction using tensor network in one dimension (1D) and numerically demonstrate that the wavefunction works well in the whole regime of $U/t$.  With the analytical argument in a generic dimension and numerical benchmark in 1D, we conjecture that the wavefunction is valid in the whole Mott insulating regime down to the zero charge gap limit on a generic lattice.

With a simple extension to finite doping, the ancilla wavefunction can describe a fractional Fermi liquid (FL*) phase which violates the Luttinger theorem without symmetry breaking.  Such a state may be a strong candidate for the mysterious pseudogap metal in hole doped cuprate\cite{zhang2020pseudogap}. Here we point out that a FL* phase can in principle exist in a narrow region of $U/t$ between the Fermi liquid and the Mott insulator phase even at half filling. Such a scenario was not well explored before mainly because the conventional approaches such as slave rotor theory can not capture it. But now we can write down a legitimate wavefunction for it from a natural extension of the wavefunction for the Mott insulator. Therefore we propose to search for such an exotic state in the future numerical studies. 

\begin{figure}[ht]
\centering
\includegraphics[width=1.0\linewidth]{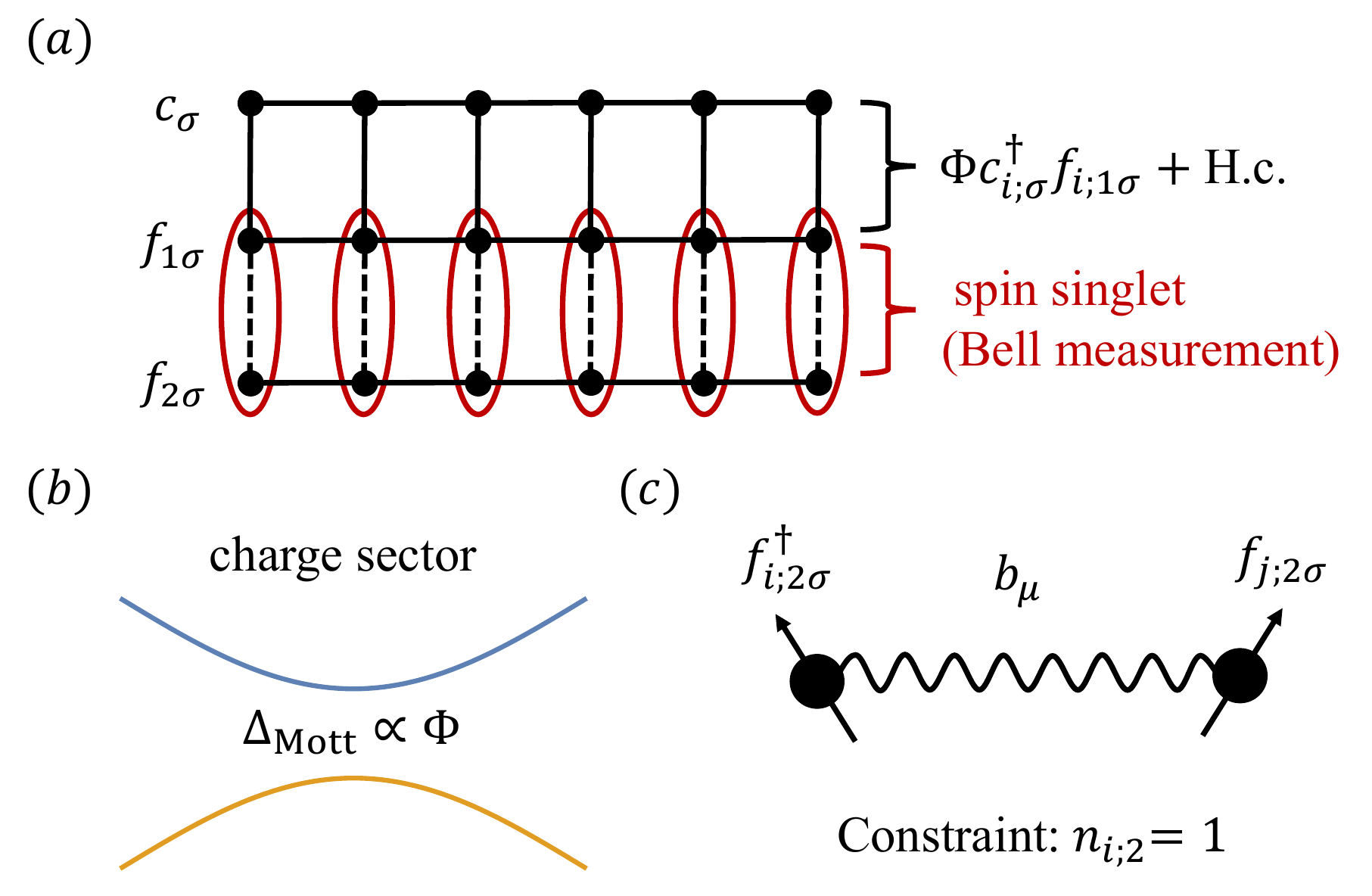}
\caption{(a) Illustration of the ancilla wavefunction formed by one physical layer and two ancilla layers. We introduce entanglement between the physical and first ancilla layer through the hybridization $\Phi$ term, and in the end project the two ancilla layers to form spin singlet at each site. The projection is equivalent to Bell measurement with post-selection and implements a quantum teleportation. (b) Upper and lower Huubard bands of the charge sector formed by $c,f_1$. (c) Spin sector formed by $f_2$. The constraint $n_{i;2}=1$ gives the $SU(2)$ gauge field $b$, the same as the familiar Abrikosov fermion theory of pure spin model\cite{wen2002quantum}.}
\label{Fig3}
\end{figure}

\textbf{Basic setup}  We consider the familiar fermionic Hubbard model:

\begin{equation}
    H=-t \sum_{\langle ij \rangle,\sigma}\left(c^\dagger_{i;\sigma} c_{j;\sigma}+\mathrm{H.c.}\right)-\mu \sum_i n_i+\frac{U}{2} \sum_i n_i (n_i-1)
\end{equation}
where $n_i=\sum_{\sigma=\uparrow,\downarrow}c^\dagger_{i;\sigma} c_{i;\sigma}$ is the density operator.  Here we only keep the nearest neighbor hopping, but the discussion below can be straightforwardly generalized to the case with longer range hopping.

Now we introduce the ancilla wavefunction. We introduce two layers of spinful ancilla fermions $f_{i;1;\tilde \sigma},f_{i;2;\tilde \sigma}$ at each site $i$. The model is shown in Fig.\ref{Fig3}(a). The ancilla wavefunction is in the form

\begin{equation}
    \ket{\Psi_c}=P \ket{\Psi_0}
\end{equation}
with $\ket{\Psi_c}$ the state in the physical Hilbert space and $\ket{\Psi_0}$ in the enlarged Hilbert space with $c$ and the ancilla $f_1,f_2$. The projection operator $P$ enforces that $f_1, f_2$ form spin singlet at each site $i$. Essentially we trace out $f_1,f_2$ to obtain a state purely in the physical Hilbert space.



In this paper we restrict to a very specific ansatz:

\begin{equation}
    \ket{\Psi_0}=\ket{\text{Slater}[c,f_1]} \ket{\text{Slater}[f_2]}
\end{equation}

Basically we assume that before projection the physical electron only couples to $f_1$, while $f_2$ forms an independent state. $\ket{\text{Slater}[c,f_1]} $ is a Slater determinant decided by the mean field Hamiltonian:  

\begin{align}\label{eqHMe}
    H^M_{e}=& - t_c \sum_{\langle ij \rangle,\sigma}\left( c^\dagger_{i;\sigma} c_{j;\sigma} +\mathrm{H.c.}\right)- \mu_c \sum_{i,\sigma} c^\dagger_{i;\sigma} c_{i;\sigma}\notag \\
    &+ t_1 \sum_{\langle ij \rangle,\sigma }\left(f^\dagger_{i;1\sigma} f_{j;1\sigma} +\mathrm{H.c.}\right)-\mu_1 \sum_{i,\sigma} f^\dagger_{i;1\sigma} f_{i;1\sigma} \notag \\ 
    &+\Phi \sum_{i,\sigma} \left(c^\dagger_{i;\sigma} f_{i;1\sigma}+\mathrm{H.c.}\right)
\end{align}
where $-\mu_1$ is introduced to fix the density of $f_1$ to be one particle per site.  $\Phi$ is the only term which entangles $c$ and the ancilla layers and is thus the most important variational parameter. Later we will see that it is proportional to the charge gap at large $U$ regime when we are at $n_c=1$ (half filling per spin).

$\ket{\text{Slater}[f_2]}$ is a Slater determinant fixed by a mean field Hamiltonian of $f_2$, which should be viewed as spinon (as will be explained below). The ansatz here can be quite general. Because $f_2$ is always fixed to be at one particle per site, one can use any  ansatz of  Gutzwiller projected wavefunction. It can be in a spin liquid state with arbitrary PSG\cite{wen2002quantum}, or one can introduce magnetic order or other symmetry breaking terms. In principle we can even use a general spin state for $f_2$ instead of a Slater determinant.  For example, one can just replace $\ket{\text{Slater}[f_2]}$ with the ground state $\ket{\Psi_s}$ of a spin model.  As we will see, the sector of $f_2$ represents the localized spin moments and has the flexibility to be in any spin state.

\begin{figure}[ht]
\centering
\includegraphics[width=1.0\linewidth]{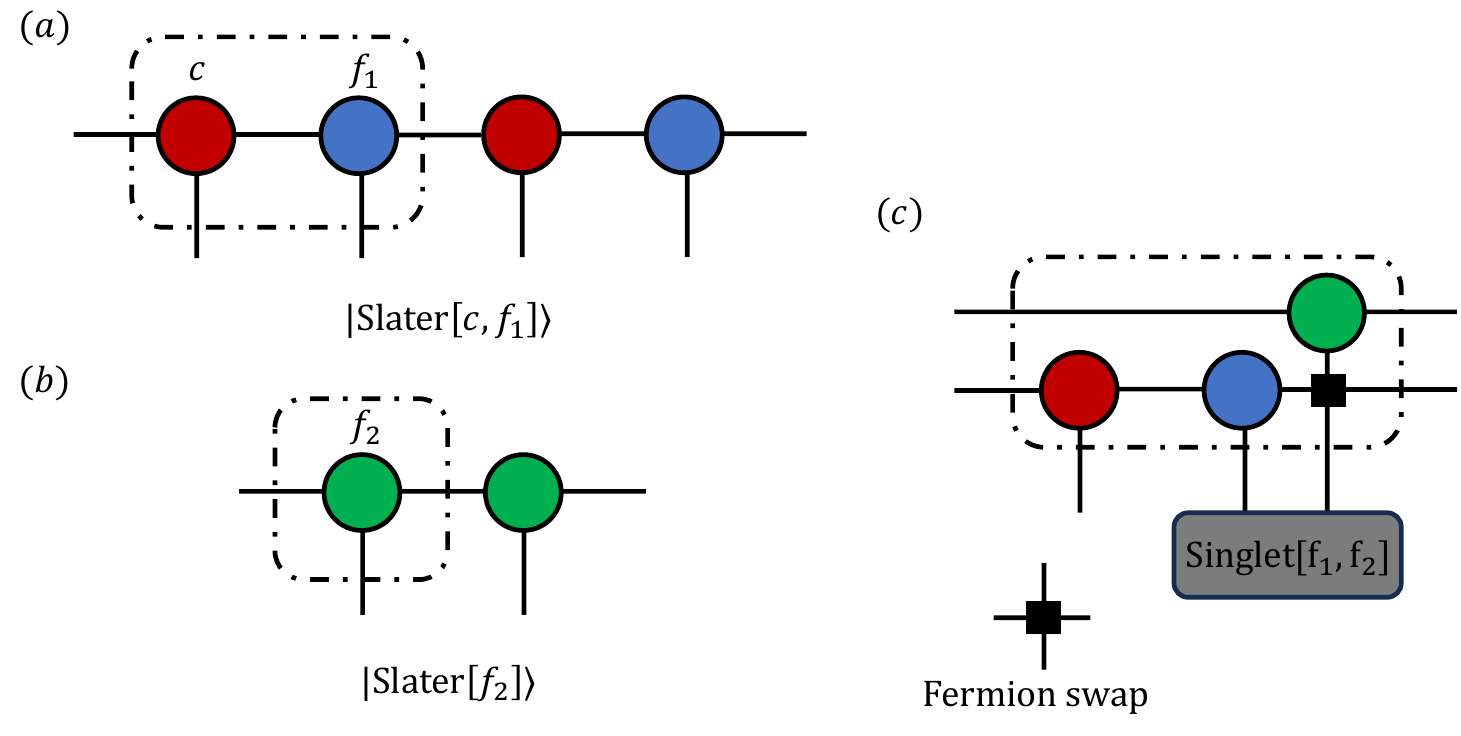}
\caption{(a) Tensor network diagram of $\ket{\mathrm{Slater}\left[c,f_1\right] }$. (b) Tensor network diagram of $\ket{\mathrm{Slater}\left[f_2\right]}$. (c) Illustration of singlet projection in one unit cell. The fermion swap is a diagonal tensor to correctly handle the fermion sign\cite{PhysRevA.81.010303,PhysRevB.81.165104}. }
\label{Fig1}
\end{figure}

Finally we project $f_1, f_2$ out and obtain a state $\ket{\Psi_c}$ purely in the physical layer.  Now $f_1, f_2$ disappear, but their influences to the physical electron remain. Conceptually it is also helpful to illustrate this wavefunction and the projection procedure using the tensor network language\cite{orus2014practical}, as shown in Fig.~\ref{Fig1}. Here we use matrix product state (MPS) in one dimension (1D) as an illustration.  A quantum state can be represented as a MPS (or projected entangled pair states (PEPS) in 2D). In the MPS, the physical leg corresponds to physical states at each site, while there are $D$ number of virtual states living on the bond.  $D$ here is the bond dimension.  In our wavefunction, we first build a MPS for $c,f_1$ (Fig.~\ref{Fig1}(a)) following $H^M_{c,f_1}$ and then build a MPS state for $f_2$ (Fig.~\ref{Fig1}(b)).  Then we combine these two tensor networks together as shown in Fig.~\ref{Fig1}(c). Now at each site the leg from the red circle corresponds to physical states, while the legs from the blue and green circles represents the states of the ancilla $f_1, f_2$. The projection corresponds to a contraction of the legs from $f_1$ and $f_2$. After the contraction, we get a tensor network with only physical legs, thus it represents a quantum state in the physical Hilbert space and can be mapped to a standad MPS for physical layer only.

\textbf{Quantum teleprotation at $\Phi=+\infty$}
We now move to study the property of  the physical state represented by the ancilla wavefunction.  First let us look at the special point with $\Phi=+\infty$ and the filling $n_c=1$. At $\Phi=+\infty$, the wavefunction of $c$ and $f_1$ can be expressed as $\ket{\mathrm{Slater}\left[c,f_1\right]}=\prod_i\left(\frac{1}{2}(c^\dagger_{i;\uparrow}-f^\dagger_{i;1\uparrow})(c^\dagger_{i;\downarrow}-f^\dagger_{i;1\downarrow})\right)\ket{0}$. At each site, $c$ and $f_1$ forms  a spin singlet, which is a perfect Einstein–Podolsky–Rosen (EPR) pair.  The projection $P$ can be understood as a Bell measurement with post selection. As in the classic quantum information context\cite{nielsen2002quantum}, the projection (Bell measurement) implements a quantum teleportaion and teleports the quantum state in $f_2$ to the physical layer $c$.  One can easily prove (see the supplementary):

\begin{equation}
    \ket{\Psi_c}=P_G \ket{\text{Slater}_{f_2}\left[c\right]}
\end{equation}
with $P_G$ as the familiar Gutzwiller projection $\prod_i (1- n_{i;\uparrow} n_{i;\downarrow})$ to enforce one electron per site.  The slater determinant ansatz is inherited from the input ansatz for $f_2$. Actually, we can also input a generic spin state for $f_2$, then $\ket{\Psi_c}$ is exactly the same as this spin state.   This is apparently a good wavefunction at infinite $U$.

\textbf{Equivalance to inverse Schrieffer-Wolff transformation at large U} In our formalism,  the constraint for the physical electron is indirectly coming from the coupling  to $f_1$ through the $\Phi$ term. When $\Phi=+\infty$, we have seen that the charge fluctuation is completely frozen in the electron layer. But because this is an indirect constraint, we can soften it simply by decreasing $\Phi$. Note in the regime with a Mott gap, we have $\mu_c=\mu_1=0$, there is freedom to change $t_c, t_1$ without changing the final wavefunction as long as $t_c+t_1$ is fixed. Here we simply fix $t_c=t, t_1=0$.  The only variational parameter is $\Phi$ in the $c,f_1$ sector. 


With a finite $\Phi$, $c$ and $f_1$ are still entangled, but they are not in a perfect EPR pair at each site. As a result, the projection only implements an imperfect quantum teleportation.  So the final state $\ket{\Psi_c}$ is the teleported spin state of $f_2$ with some corrections. At linear order of $t/U$, we can work out the correction analytically:

\begin{equation}
    \ket{\Psi_c(\Phi=\frac{U}{2})}\approx e^{-\mathrm{i} S} \ket{\Psi_c(\Phi=+\infty)}
\end{equation}
where $e^{-\mathrm{i}S}$ is the inverse of the Schrieffer-Wolff transformation in the standard $t/U$ expansion of the Hubbard model\cite{macdonald1988t}.  Therefore, at large $U$ regime, our ancilla wavefunction  manages to capture the charge fluctuation correctly with $\Phi=\frac{U}{2}$. At large $U$ we do not even need to optimize $\Phi$.  The only free parameters are in the spin state of the $f_2$ sector, but the various ansatz of the spin states have been well studied and we can just employ them in a specific model.  As long as one can find a good wavefunction for the spin state, our ancilla approach can obtain a good wavefunction with finite charge gap. In the supplementary we prove that the final wavefunction is symmetric even if the spin state of $f_2$ has a non-trivial PSG, in contrast to the conventional Jastrow factor approach.

\textbf{Physical meaning of the ancilla fermions} In the example of $\Phi=+\infty$ and large $\Phi$, we can already see that the state of $f_2$ becomes the spin state of physical layer after projection.  Let us provide a more intuitive interpretation of the ancilla $f_1, f_2$.  Our wavefunction has the structure $\ket{\Psi_c}=P \ket{\Psi_0}$. Suppose $\ket{\Psi_0}$ is a good ansatz such that $\ket{\Psi_c}$ is a good variational ground state. We can then apply $f_1, f_2$ operators to $\ket{\Psi_0}$ to obtain variational excited states. When $\Phi=+\infty$, we can prove (see the supplementary) that 

\begin{equation}
P f^\dagger_{i;1\sigma} \ket{\Psi_0}=c^\dagger_{i;\sigma} \ket{\Psi_c},
\end{equation}

\begin{equation}
P f^\dagger_{i;2\sigma} \ket{\Psi_0}=0,
\end{equation}

and
\begin{equation}
P f^\dagger_{i;2\sigma} f_{i;2\sigma'}\ket{\Psi_0}=c^\dagger_{i;\sigma}c_{i;\sigma'} \ket{\Psi_c}.
\end{equation}

We can see that the action of $f_1^\dagger$ is the same as an electron operator while $f_2^\dagger $ alone is meaningless.  $f_{i;2\sigma}^\dagger f_{i;2\sigma}$ however is meaningful and behaves as a physical spin operator. One natural interpretation is that $f_1$ is the same as the electron operator while $f_2$ is a neutral spinon operator when we have a large $\Phi$.

A better way is to use gauge theory to implement the projection.  Our projection $P$ corresponds to three constraint at each site $i$: (I) $n_{i;1}=1$; (II) $n_{i;2}=1$; (III) $\vec S_{i;1}+\vec S_{i;2}=0$ where $\vec S_{i;a}$ is the spin operator of the ancilla $f_{a}$, $a=1,2$. The first two constraints are similar to the usual Gutzwiller projection constraints and therefore can be fixed with U(1) gauge field (SU(2) in the case of spin $1/2$ fermion due to additional particle-hole transformation\cite{wen2002quantum}). The third condition constrains $f_1, f_2$ to form spin singlet and introduces a SU(2) gauge field\cite{zhang2020deconfined}.  Let us label the three SU(2) gauge field from the three constraints  as $a$, $b$ and $\alpha$ respectively. Then $f_1$ couples to $a$ and $\alpha$ while $f_2$ couples to $b$ and $\alpha$. Physical electron $c$ is gauge invariant as expected.   In our ansatz with a term $-\Phi \sum_{i,\sigma}\left( c^\dagger_{i;\sigma}f_{i;1\sigma}+\mathrm{H.c.}\right)$,  $a$ and $\alpha$ are both higgsed as they couple to $f_1$\cite{zhang2020deconfined}. Hence we are in the higgsed phase of the gauge theory. Now we can ignore $a$ and $\alpha$. $f_1$ does not couple to any gauge field and can be identified as the electron\footnote{More precisely, the gauge field $a$ is locked to the external probing field $A$, then $f_1$ also couples to the physical electric and magnetic field as an electron.}. In contrast, the gauge field $b$ is untouched by the term $\Phi$ and $f_2$ still couples to it unless the ansatz of $f_2$ completely higgses $b$. The coupling of $f_2$ to $b$ is basically the same as the familiar Abrikosov fermion description of spin liquid, so $f_{2\sigma}$ can be identified exactly as the usual Abrikosov fermionic spinon\cite{wen2002quantum} and we can recover any spin liquid states.   Especially, even if $f_2$ is in an ansatz with projective symmetry group, the final state is still symmetric.

In summary, with a finite $\Phi$, we have two separate sectors: a charge sector formed by $c,f_1$ and a spin sector formed by the spinon $f_2$ as shown in Fig.\ref{Fig3}(b) and (c). In the charge sector, we have two bands which can be interpreted as the upper Hubbard and lower Hubbard band with the charge gap $\Delta_c=2\Phi$. The spin sector allows any spin states. Therefore our formalism has an explicit spin charge separation which is expected for Mott insulator.It is naturally to expect that the wavefunction can extend further to the regime with very small charge gap simply by decreasing $\Phi$.  We will confirm this conjecture for one dimension through explicit numerical simulation.

\textbf{Numerical benchmark in 1D} In 1D, We construct the ancilla wavefunction via tensor network using the TeNPy Library (version 0.9.0)\cite{tenpy}.  Initially, we construct two fermionic gaussian states using correlation matrices\cite{PhysRevB.92.075132} for $c$ and $f_1$  in the spin up and down channel respectively. Subsequently, we employed the same method to construct the state $|\mathrm{Slater}[f_2]\rangle$. Finally, we employed the Gutzwiller zipper method \cite{aghaei2020efficient} to perform a tensor product of these three states and implement the desired projection.

\begin{figure}[ht]
\centering
\includegraphics[width=1.0\linewidth]{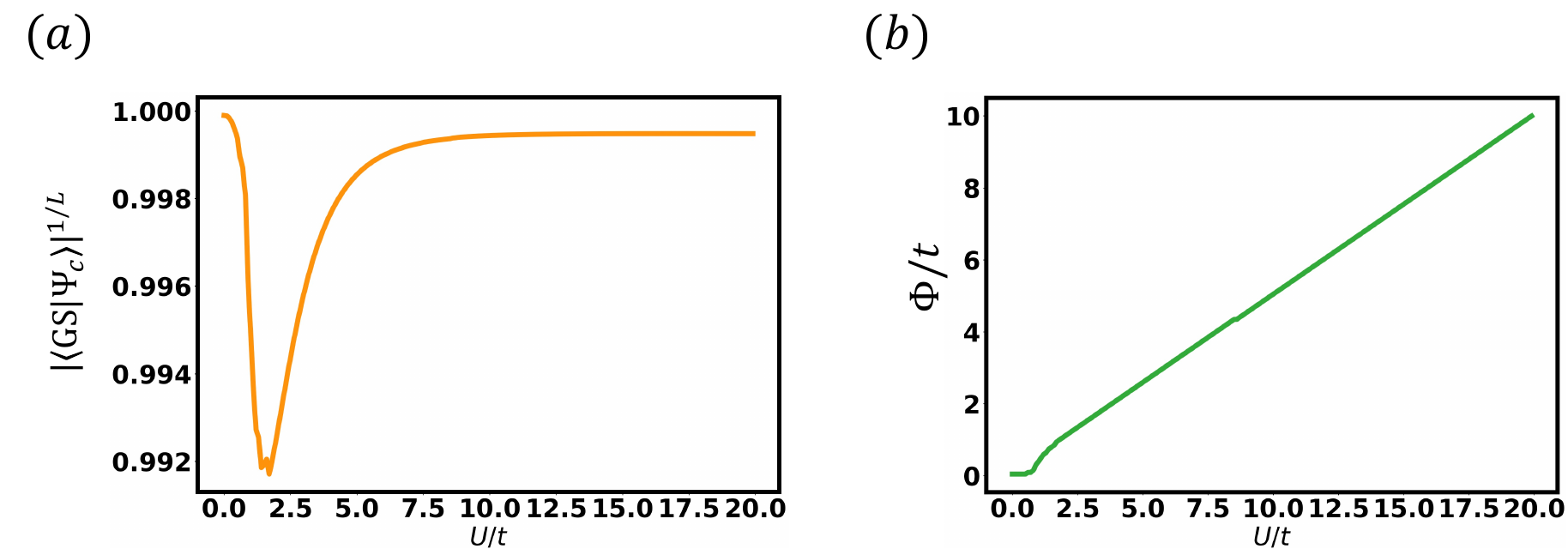}
\caption{Benchmark of the ancilla wavefunction with the DMRG results at $L=50$. (a) The overlap of the ancilla wavefunction and the Ground State as computed via DMRG with $U/t$. (b) The  optimized parameter $\Phi/t$ with $U/t$. One find $\Phi \sim U/2$ as expected from the general analytical arguments above.}
\label{Fig4}
\end{figure}

Our ansatz of $f_2$ has only nearest neighbor hopping, which is known to work well in 1D. At each $U/t$, we optimize $\Phi$ to minimize the energy of the Hubbard model.  To assess the quality of our wavefunction $\ket{\Psi_c}$, we calculate the overlap with the ground state $\ket{\mathrm{GS}}$ obtained by the Density Matrix Renormalization Group (DMRG) method. The results are presented in Fig.\ref{Fig4}. Notably, the overlap per unit cell is larger than $0.99$ in the whole range of $U/t$, indicating the good performance of the ancilla wavefunction with one single variational parameter $\Phi$.

\textbf{Intermediate pseudogap metal} We have demonstrated the power of the ancilla wavefunction in the Mott insulating regime. The wavefunction can be easily extended to finite doping by tuning $\mu_c$ for electron, which usually leads to a pseudogap metal with small Fermi surface. If $f_2$ is in a spin liquid phase, this describes a fractional Fermi liquid (FL*) which violates the Luttinger theorem but does not break any symmetry. This approach has been demonstrated to be attractive phenomenologically for underdoped cuprates\cite{zhang2020pseudogap,mascot2022electronic,christos2023model}. Here wee point out that a similar FL* phase may be possible around the metal-insulator transition even at $n_c=1$, the undoped sample.

\begin{figure}[ht]
\centering
\includegraphics[width=1.0\linewidth]{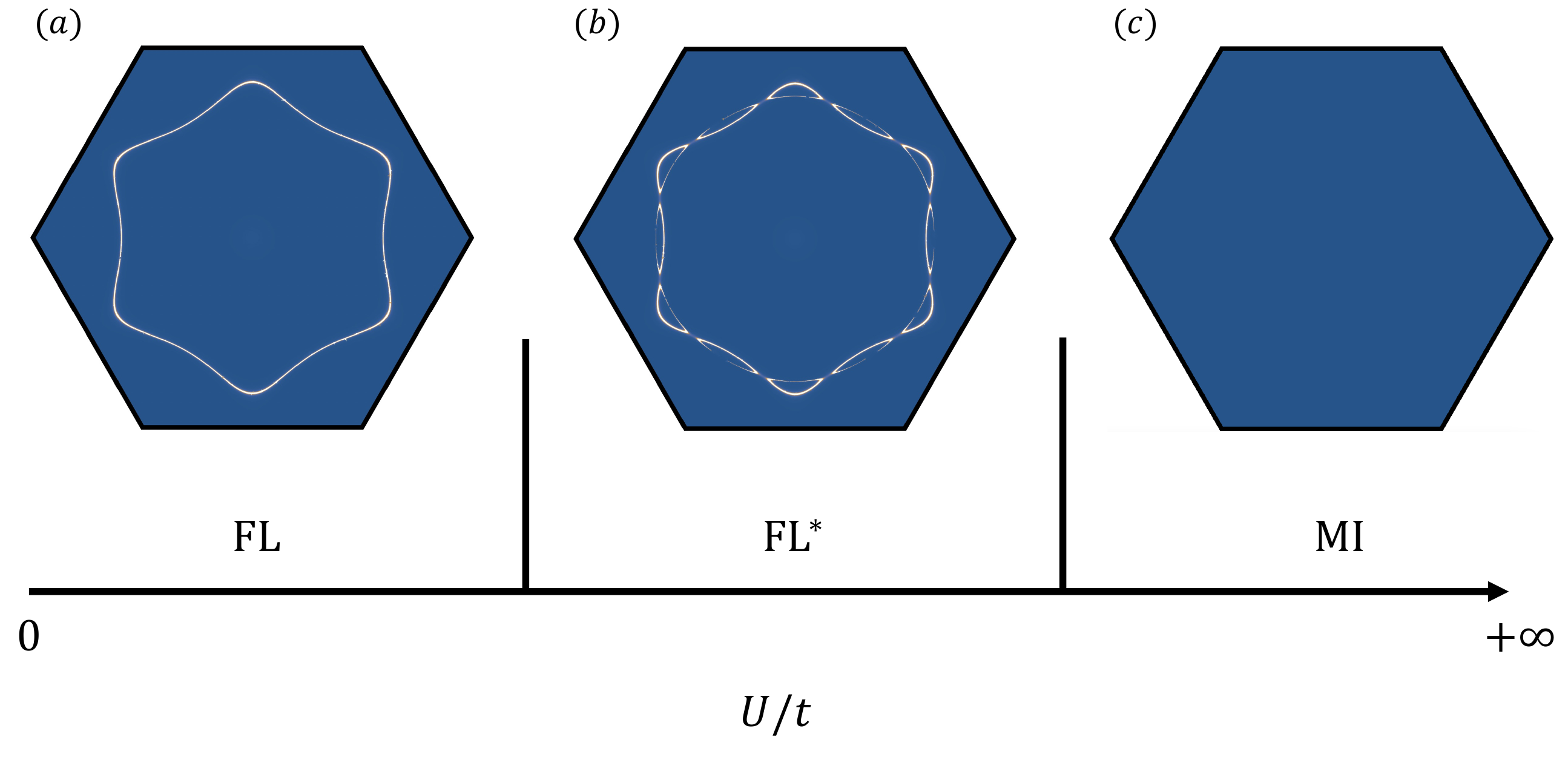}
\caption{A conjectured phase diagram tuned by $U/t$ at $n_c=1$ on triangular lattice. The left critical point is associated with the onset of $\Phi$, while the right critical point is simply a Lifshitz transition.  FL, FL* and MI label Fermi liquid, fractional Fermi liquid and Mott insulator. Across the two transitions we assume $f_2$ in a chiral spin liquid ansatz (see the supplementary).  In layer $c$, both nearest neighbor (NN) and next nearest neighbor (NNN) hoppings exist, with respective values of $t_c=1$ and $t_c'=-0.82$. In layer $f_1$, we assume only NN hopping with a value of $t_1=1$.$(a)-(c)$ are the color plots of the electron spectral weight $A_c(\mathbf{k},\omega=0)$.  In (a), $\Phi=0$, we have a Fermi liquid. In (b), $\Phi=0.1$, we have a FL* phase with equal size of electron and hole pockets and dominated spectral weight in disconnected fermi arcs. In (c), we are in a Mott insulator phase. }
\label{Fig5}
\end{figure}

At $n_c=1$, we have shown that $\Phi\sim U/2$ in the large U regime and describes a Mott insulator in our ancilla wavefunction.  The Fermi liquid corresponds to $\Phi=0$. Then naturally one may expect the metal-insulator transition tuned by $U/t$ is associated with the onset of $\Phi$ in the ancilla wavefunction. However, generically the fermi surfaces from $c$ and $f_1$ may have a small mismatch (unless in 1D) without fine tuning. Then a small $\Phi$ can not fully gap out all the Fermi surface, leading to a pseudogap metal. On triangular lattice, there is evidence that the weak Mott insualtor is in a chiral spin liquid(CSL), so we should keep $f_2$ in a CSL ansatz (see the supplementary for details).  Then there can be an intermeidate FL* phase as illustrated in Fig.~\ref{Fig5}. One can see equal size of electron and hole pockets where the backside is from $f_1$ and thus has small spectral weight. Therefore at finite temperature one may observe disconnected fermi arcs as in the underdoped cuprate. As $\Phi$ increases with increasing $U/t$, these pockets shrink and eventually evolve into a Mott insulator. Right now we do not know the energetics of this intermeidate FL* phase, but at least we can write a legitimate wavefunction. We hope to do serious numerical study to search for this exotic scenario in future.

\textbf{Conclusion} In summary, we benchmark  the recently proposed ancilla wavefunction\cite{zhang2020pseudogap} for Mott insulator with finite charge gap. Analytically we show that the wavefunction reduces to the Gutzwiller wavefunction in the $U=\infty$ limit at the filling $n=1$. At leading order of $t/U$, the wavefunction is equivalent to inverse Schriffer-wolffe.  We then conjecture that the wavefunction also works well in the weak Mott regime. We bench mark the wavefunction in 1D Hubbard model at $n=1$ using a tensor network representation. We also argue that a FL* phase may exist in the intermediate U regime of a Hubbard model at filling $n=1$. Given the flexibility of the wavefunction to capture any spin states including spin liquids with non-trival PSG, we anticipate the ancilla wavefunction will be a useful numerical tool and theoretical framework in dealing with the weak Mott regime and pseudogap metal.

\textbf{Acknowledgement} We thank Subir Sachdev and Henry Shackleton for discussions. We thank Zhehao Dai for discussions on tensor network algorithm.  This work was supported by the
National Science Foundation under Grant No. DMR-2237031. The work by YHZ was  performed in part at Aspen Center for Physics, which is supported by National Science Foundation grant PHY-2210452. The numerical simulation was  carried out at the
Advanced Research Computing at Hopkins (ARCH) core facility (rockfish.jhu.edu), which is supported by the National
Science Foundation (NSF) grant number OAC 1920103.

\bibliographystyle{apsrev4-1}
\bibliography{main}

\appendix 

\onecolumngrid

\section{Jastrow and ancilla wavefunction at finite U with non-trivial PSG}

In this section we compare the familiar Jastrow wavefunction with our ancilla wavefunction in the regime with a finite U.  We will argue that the Jastrow factor approach fails to represent any spin liquid state with a non-trivial projective symmetry group (PSG).

A general form of the Jastrow factor wavefunction is:

\begin{equation}
    \ket{\Psi_{J}}=e^{- \frac{1}{2} \sum_{ij}V_{ij} n_i n_j} |\text{Slater}[c]\rangle
    \label{eq:jastrow}
\end{equation}

One first constructs a slater Determinant following a mean field ansatz, and then add a Jastrow factor to suppress the charge fluctuations.   If one uses the Gutzwiller projection $P_G=\prod_i (1-n_{i;\uparrow}n_{i;\downarrow})$, then this state is a pure spin state without charge fluctuation. In this case the fermion in the Slater determinant should be interpreted as a neutral spinon. However, if we use the Jastrow factor instead, the fermion in the Slater determinant now carries both spin and charge.  With a sufficiently singular $V_{ij}$ we can open a charge gap and describes a Mott insulator with spin liquid ansatz, but now the density-density correlation is influenced by the spin liquid ansatz.   On the other hand, in a Mott insulator we should expect spin charge separation:  the suppression of the density fluctuation happens at energy scale of $\Delta_c \sim U$, while the spin liquid part is decided by the energy scale of $J \sim \frac{4 t^2}{U}\ll\Delta_c$. Therefore we should expect that the density fluctuations is roughly independent of the spin liquid ansatz. The Jastrow factor wavefunction apparently fails on this aspect.

\begin{figure}[ht]
\centering
\includegraphics[width=0.5\linewidth]{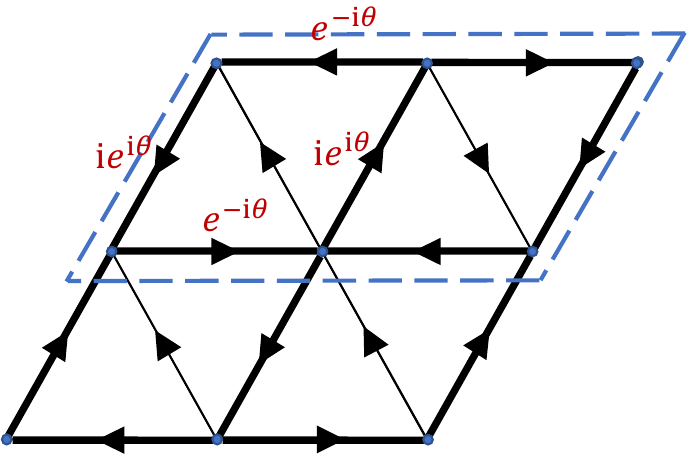}
\caption{A chiral spin liquid ansatz on triangular lattice for fermionic spinons. The blue dashed line indicates the doubled unit cell. The orientation of the arrow along the bond represents the phase of the hopping parameter in the ansatz and there is a flux $\pi$ per unit cell. In the Jastrow factor approach, the charge fluctuation will fell the doubled unit cell and the final state breaks translation symmetry. In contrast, in the ancilla wavefunction, the ancilla $f_2$ is put in this ansatz, but the final state is symmetric because the density fluctuation of $f_2$ is completely frozen as in the familiar Gutzwiller projected wavefunction.}
\label{fig:csl_ansatz}
\end{figure}

To further reveal the problem, we consider a chiral spin liquid ansatz on triangular lattice.   We have fermionic spinons in an ansatz with doubled unit cell and a flux $\pi$ per unit cell, as illustrated in Fig.~\ref{fig:csl_ansatz}. Now the translation symmetry  is broken in the mean field level.  In the Gutzwiller projected wavefunction, the resulting state is still translation invariant because the spinon can have a projective symmetry group and there is no charge fluctuation.

But at finite U, in the Jastrow factor wavefunction,  we need to put $\ket{\text{Slater}[c]}$ in this ansatz with doubled unit cell. We can then  calculate the expectation value of the bond current operator

\begin{equation}
    \langle T_{ij} \rangle=\sum_{\sigma}\langle ic^\dagger_{i\sigma} c_{j\sigma} -i c^\dagger_{j\sigma} c_{i\sigma} \rangle
\end{equation}

As long as $U$ is finite, the Jastrow factor wavefucntion in Eq.~\ref{eq:jastrow}  allows density fluctuations and we expect that $\langle T_{ij} \rangle \neq 0$ and follows the same pattern as the mean field ansatz in Fig.~\ref{fig:csl_ansatz}. Therefore $\langle T_{ij} \rangle$  breaks translation symmetry.  The Jastrow factor suppresses the density fluctuations, but can not alter this property qualitatively. As a result, the final wavefunction is not translationally invariant and thus is a quite poor ansatz for a chiral spin liquid.   The problem of the Jastrow wavefunction orginates from the artifact that the charge and spin are represented by the same fermion, while in a Mott insulator we expect seperation of the charge sector and spin sector.

On the other hand, our ancilla wavefunction can easily represent a translationally invariant chiral spin liquid with arbitrary charge gap. We simply put our ancilla $f_2$ in the mean field ansatz shown in Fig.~\ref{fig:csl_ansatz}. Our final projection $P$ enforces that there is only one $f_2$ particle per site, the same as the familiar Gutzwiller projector at infinite U for the spin model. So there is no density fluctuations for $f_2$ and the final wavefunction is translation invariant.  Note here $f_2$ is purely neutral and is the same as the Abrikosov fermion, so its mean field ansatz allows non-trivial PSG\cite{wen2002quantum}.  Meanwhile the charge sector is separately represented by $c,f_1$ and the charge gap is controlled by $\Phi$. Clealy there is no strict requirement of the spin ansatz or PSG at small charge gap.

\section{Ancilla wavefunction of Mott insulator at $\Phi=+\infty$}

In the limit $\Phi=+\infty$, the hopping  terms of $c$ and $f_1$ can be ignored. Therefore, $c$ and $f_1$ form spin-singlet at each site: 
\begin{equation}\label{eqGS1}
    \ket{\mathrm{Slater}[c,f_1]}=\prod_{i=1}^N\left(\frac{1}{2}(c^\dagger_{i;\uparrow}-f^\dagger_{i;1\uparrow})(c^\dagger_{i;\downarrow}-f^\dagger_{i;1\downarrow})\right)\ket{0},
\end{equation}
in which $i$ is the site index. Here we assume a generic lattice on arbitrary dimension. The wavefunction of $f_2$ can be written in the form:
\begin{equation}\label{eqGS2}
    \ket{\Psi_2}=P_G\ket{\mathrm{Slater}[f_2]}=\sum_{\{\sigma_i\}}\psi(\{\sigma_i\}) \prod_{i=1}^Nf^\dagger_{i;2\sigma_i}\ket{0}.
\end{equation}
In the main text we define that $\ket{\Psi_0}=\ket{\mathrm{Slater}[c,f_1]}\ket{\mathrm{Slater}[f_2]}$, here for convenience we change it to be:
\begin{equation}\label{eqGS}
\ket{\Psi_0}=\ket{\mathrm{Slater}[c,f_1]}\ket{\Psi_2},
\end{equation}
which still satisfies $\ket{\Psi_c}=P\ket{\Psi_0}=\sum_c\langle c,s|\Psi_0\rangle|c\rangle$. Here $\ket{c}$ is summed over the many body basis of the physical Hilbert space.

We combine Eq.\ref{eqGS1}, Eq.\ref{eqGS2} and Eq.\ref{eqGS}, the final expression of $\ket{\Psi_0}$ is written as:
\begin{equation}\label{eqGSform}
\begin{split}
    \ket{\Psi_0}=&\sum_{\{\sigma_i\}}\psi(\{\sigma_i\})\prod_{i=1}^N\left(\frac{1}{2}(c^\dagger_{i;\uparrow}-f^\dagger_{i;1\uparrow})(c^\dagger_{i;\downarrow}-f^\dagger_{i;1\downarrow})f^\dagger_{i;2\sigma_i}\right)\ket{0}\\
    =&\sum_{\{\sigma_i\}}\psi(\{\sigma_i\})\prod_{i=1}^N\left(\frac{1}{2}(-\sigma_ic^\dagger_{i;\sigma_i}f^\dagger_{i;1\bar{\sigma}_i}f^\dagger_{i;2\sigma_i}+\mathrm{other \ terms})\right)\ket{0}.
\end{split}
\end{equation}
where $\sigma_i=1,-1$ for spin up and down respectively. The projection operator $P=\sum_c |c\rangle\langle c,s|$ enforces $f_1$ and $f_2$ form spin singlet at each site, it can be written as:


\begin{equation}
   P=\prod_{i=1}^N \left(\frac{1}{\sqrt{2}}(1-n_{i;1\uparrow})(1-n_{i;1\downarrow})(f_{i;2\downarrow}f_{i;1\uparrow}-f_{i;2\uparrow}f_{i;1\downarrow})\right)=\prod_{i=1}^N P_i,
\end{equation}
where we define $P_i=\frac{1}{\sqrt{2}}(1-n_{i;1\uparrow})(1-n_{i;1\downarrow})(f_{i;2\downarrow}f_{i;1\uparrow}-f_{i;2\uparrow}f_{i;1\downarrow})$. Substituting into Eq.\ref{eqGSform} and multiplying a normalization factor, we obtain the final state:
\begin{equation}\label{eqGSphiinfty}
    \ket{\Psi_c}=\sum_{\{\sigma_i\}}\psi(\{\sigma_i\})\prod_{i=1}^N c^\dagger_{i;\sigma_i}\ket{0},
\end{equation}
which is the same as Eq.\ref{eqGS2}.

We can take a closer look at Eq.\ref{eqGSform} and how operators affect the final projected state. The basic two cases are simply applying one $c/c^\dagger$ or one $f_1/f_1^\dagger$ operator on the ground state before doing the projection. For the operator on the physical layer, it is obvious that:
\begin{equation}
    c_{j;\sigma}\ket{\Psi_0}=\sum_{\{\sigma_i\}}\psi(\{\sigma_i\})(-1)^{\sum_{k<j}n_k}\prod_{i=1}^N\left(\frac{1}{2}(\delta_{ij}c_{j;\sigma}+1-\delta_{ij})(c^\dagger_{i;\uparrow}-f^\dagger_{i;1\uparrow})(c^\dagger_{i;\downarrow}-f^\dagger_{i;1\downarrow})f^\dagger_{i;2\sigma_i}\right)\ket{0},
\end{equation}
then we have,
\begin{equation}
    \begin{split}
        Pc_{j;\sigma}\ket{\Psi_0}=&(-1)^{j-1}\sum_{\{\sigma_i\}}\psi(\{\sigma_i\})\prod_{i=1}^N (\delta_{ij}c_{j;\sigma}+1-\delta_{ij})c^\dagger_{i;\sigma_i}\ket{0}\\
        =&c_{j;\sigma}\ket{\Psi_c}=c_{j;\sigma}P\ket{\Psi_0}.
    \end{split}
\end{equation}
As for $f_1$ operator, we have that:
\begin{equation}
\begin{split}
    f_{j;1\sigma}\ket{\Psi_0}=&\sum_{\{\sigma_i\}}\psi(\{\sigma_i\})(-1)^{\sum_{k<j}n_{k}}\prod_{i=1}^N\left(\frac{1}{2}(\delta_{ij}f_{j;1\sigma}+1-\delta_{ij})(c^\dagger_{i;\uparrow}-f^\dagger_{i;1\uparrow})(c^\dagger_{i;\downarrow}-f^\dagger_{i;1\downarrow})f^\dagger_{i;2\sigma_i}\right)\ket{0}\\
    =&\sum_{\{\sigma_i\}}\psi(\{\sigma_i\})(-1)^{\sum_{k<j}n_{k}}\prod_{i=1}^N\frac{1}{2}\left(\left(-(1-\delta_{ij})c^\dagger_{i;\sigma_i}+\delta_{ij}\delta_{\sigma\sigma_i}\right)\sigma_if^\dagger_{i;1\bar{\sigma}_i}f^\dagger_{i;2\sigma_i}+\mathrm{other \ terms}\right)\ket{0}.
\end{split}
\end{equation}
After the projection and the renormalization we can get:
\begin{equation}
\begin{split}
    Pf_{j;1\sigma}\ket{\Psi_0}=&\sum_{\{\sigma_i\}}\psi(\{\sigma_i\})(-1)^{j-1}\prod_{i=1}^N\left((1-\delta_{ij})c^\dagger_{i;\sigma_i}-\delta_{ij}\delta_{\sigma\sigma_i}\right)\ket{0}\\
    =&-c_{j;\sigma}\ket{\Psi_c}.
\end{split}
\end{equation}
By doing the same procedure, we can also prove that:
\begin{equation}
    \begin{split}
        Pc^\dagger_{j;\sigma}\ket{\Psi_0}=&c^\dagger_{j;\sigma}\ket{\Psi_c},\\
        Pf^\dagger_{j;1\sigma}\ket{\Psi_0}=&c^\dagger_{j;\sigma}\ket{\Psi_c}.
    \end{split}
\end{equation}
If we write $P$ as $\prod_{i=N}^1 P_i$, the above results can be wtitten as:
\begin{equation}
    \begin{split}
        P_j c_{j;\sigma}\ket{\Psi_0}=&c_{j;\sigma} P_j\ket{\Psi_0},\\
        P_j c^\dagger_{j;\sigma}\ket{\Psi_0}=&c^\dagger_{j;\sigma} P_j\ket{\Psi_0},\\
        P_j f_{j;1\sigma}\ket{\Psi_0}=&-f_{j;1\sigma} P_j\ket{\Psi_0},\\
        P_j f^\dagger_{j;1\sigma}\ket{\Psi_0}=&f^\dagger_{j;1\sigma} P_j\ket{\Psi_0}.
    \end{split}
\end{equation}
For the case with more than one operator, we can sort the order with the site index increasing:
\begin{equation}
    O=(-1)^F\prod_{i=1}^N O_i,
\end{equation}
where $(-1)^F$ comes from the exchange of fermionic operators. The projected state is:
\begin{equation}
\begin{split}
   P O \ket{\Psi_0}=&(-1)^F (P_NP_{N-1}...P_1)(O_1O_2...O_N)\ket{\Psi_0}\\
   =&(-1)^F (P_1O_1)(P_2O_2)...(P_NO_N)\ket{\Psi_0}.
\end{split}
\end{equation}
Since we act as $(P_1O_1)(P_2O_2)...(P_NO_N)$, and the $\ket{\mathrm{GS}}$ is written in Eq.\ref{eqGSform} with site index increasing, so the effect of the operator $P_iO_i$ would not be affected by $(P_{i+1}O_{i+1})(P_{i+2}O_{i+2})...(P_NO_N)$. For example, we still have that:
\begin{equation}
    P_i c_{i;\sigma}(P_{i+1}O_{i+1})(P_{i+2}O_{i+2})...(P_NO_N)\ket{\Psi_0}=c_{i;\sigma}P_i(P_{i+1}O_{i+1})(P_{i+2}O_{i+2})...(P_NO_N)\ket{\Psi_0}.
\end{equation}
Therefore, we can prove the following relation:
\begin{equation}\label{eqcdf}
\begin{split}
    Pc^\dagger_{i;\sigma}f_{j;1\sigma}\ket{\Psi_0}=&-c^\dagger_{i;\sigma}c_{j;1\sigma}\ket{\Psi_c},\\
    Pf^\dagger_{i;1\sigma}c_{j;\sigma}\ket{\Psi_0}=&c^\dagger_{i;1\sigma}c_{j;\sigma}\ket{\Psi_c}.
\end{split}
\end{equation}

As for $f_2$ operator, $f_{i;2\sigma}/f_{i;2\sigma}^\dagger$ changes the $n_{i;2}$, therefore we have:
\begin{equation}
    \begin{split}
        P_if_{i;2\sigma}\ket{\Psi_0}=&0,\\
        P_if^\dagger_{i;2\sigma}\ket{\Psi_0}=&0.
    \end{split}
\end{equation}
For the case with two operators on the same site $i$, we have:
\begin{equation}
\begin{split}
    Pf^\dagger_{i;2\sigma}f_{i;2\sigma^\prime}\ket{\Psi_0}=&P\ket{\mathrm{Slater}[c,f_1]}f^\dagger_{i;2\sigma}f_{i;2\sigma^\prime}\ket{\Psi_2}\\
=&P\ket{\mathrm{Slater}[c,f_1]}f^\dagger_{i;2\sigma}f_{i;2\sigma^\prime}\sum_{\{\sigma_j\}}\psi(\{\sigma_j\}) \prod_{j=1}^Nf^\dagger_{j;2\sigma_j}\ket{0},
\end{split}
\end{equation}
since $f^\dagger_{i;2\sigma}f_{i;2\sigma^\prime}$ only changes $\psi(\{\sigma_j\})$, by Eq.\ref{eqGSphiinfty} we can conclude that:
\begin{equation}
    Pf^\dagger_{i;2\sigma}f_{i;2\sigma^\prime}\ket{\Psi_0}=c^\dagger_{i;\sigma}c_{i;\sigma^\prime}\ket{\Psi_c}.
\end{equation}

\section{Ancilla wavefunction in Hubbard model at large U}
In this section, we are going to use the inverse of the Schrieffer-Wolff transformation to represent the ground state of Hubbard model at large $U$ and $n=1$, then prove that it is equivalent to our ancilla wavefunction. First, the generic fermionic Hubbard model can be written as:
\begin{equation}\label{eqHubbard}
    H_{\mathrm{Hubbard}}=-\sum_{i,j,\sigma}t_{ij}c^\dagger_{i;\sigma}c_{j;\sigma}+U\sum_i n_{i;\uparrow}n_{i;\downarrow}=T+V,
\end{equation}
From Ref.\cite{macdonald1988t} we know that the kinetic part can be written as:
\begin{equation}
    \begin{split}
        T=&T_0+T_1+T_{-1},\\
        T_0=&-\sum_{i,j,\sigma}t_{ij}\left(n_{i;\bar{\sigma}}c^\dagger_{i;\sigma}c_{j;\sigma}n_{j;\bar{\sigma}}+h_{i;\bar{\sigma}}c^\dagger_{i;\sigma}c_{j;\sigma}h_{j;\bar{\sigma}}\right),\\
        T_1=&-\sum_{i,j,\sigma}t_{ij}n_{i;\bar{\sigma}}c^\dagger_{i;\sigma}c_{j;\sigma}h_{j;\bar{\sigma}},\\
        T_{-1}=&-\sum_{i,j,\sigma}t_{ij}h_{i;\bar{\sigma}}c^\dagger_{i;\sigma}c_{j;\sigma}n_{j;\bar{\sigma}},
    \end{split}
\end{equation}
where $h_{i;\sigma}=1-n_{i;\sigma}$, $T_0$ doesn't change the number of the double occupied sites, $T_1$ and $T_{-1}$ increases and decreases the number of the double occupied sites respectively.
In the large $U$ limit, the double/zero occupancy at each site is forbidden, the Hamiltonian becomes \cite{macdonald1988t}:
\begin{equation}\label{eqH0}
    H_{\mathrm{Hubbard},0}=T_0+V+U^{-1}\left([T_1,T_{-1}]+[T_0,T_{-1}]+[T_1,T_0]\right).
\end{equation}
We can perform a unitary transformation on the above Hamiltonian and obtain that \cite{macdonald1988t}:
\begin{equation}
        H^\prime=e^{-\mathrm{i}S} H_{\mathrm{Hubbard},0} e^{\mathrm{i}S}=V+T_0+T_1+T_{-1}+O(U^{-2}),
\end{equation}
in which $\mathrm{i}S=U^{-1}(T_1-T_{-1})$. $H^\prime$ is the same as $H_{\mathrm{Hubbard}}$ to $O(U^{-1})$ order. The ground state of the above Hamiltonian can be obtained by performing the unitary transformation $e^{-\mathrm{i}S}$ on the ground state of Eq.\ref{eqH0}, which is a pure spin wavefunction, we assume it to be Eq.\ref{eqGSphiinfty} as argued in the main text. The ground state of Eq.\ref{eqHubbard} can be written as:
\begin{equation}\label{eqGSHubbard}
\begin{split}
    \ket{\mathrm{GS},\mathrm{Hubbard}}=&e^{-\mathrm{i}S}\sum_{\{\sigma_i\}}\psi(\{\sigma_i\})\prod_{i=1}^N c^\dagger_{i;\sigma_i}\ket{0}\\
    \approx&\left(1-\mathrm{i}S\right)\sum_{\{\sigma_i\}}\psi(\{\sigma_i\})\prod_{i=1}^N c^\dagger_{i;\sigma_i}\ket{0}.
\end{split}
\end{equation}
Since there is no double occupancy and zero occupancy for Eq.\ref{eqGSphiinfty}, so $T_{-1}=0$ and $T_1=-t\sum_{\braket{ij},\sigma}(c^\dagger_{i;\sigma}c_{j;\sigma}+\mathrm{H.c.})$ automatically. Therefore, Eq.\ref{eqGSHubbard} can be rewritten as:
\begin{equation}\label{eqGSUfinite}
    \ket{\mathrm{GS},\mathrm{Hubbard}}=\left(1+\sum_{i,j,\sigma}\frac{t_{ij}}{U}c^\dagger_{i;\sigma}c_{j;\sigma}\right)\sum_{\{\sigma_i\}}\psi(\{\sigma_i\})\prod_{i=1}^N c^\dagger_{i;\sigma_i}\ket{0}.
\end{equation}

The above equation can be represented as the ancilla wavefunction at large $\Phi$. The hopping ansatz of $c$ and $f_1$ is represented as $t_{c,ij}$ and $t_{1,ij}$ respectively. The fourier transform of them are $h_c(\mathbf{k})$ and $h_{f_1}(\mathbf{k})$. Then the mean field Hamiltonian of $c$ and $f_1$ can be written as:
\begin{equation}
\begin{split}
H^M_{e}=&\sum_{\mathbf{k},\sigma} \begin{pmatrix}
    c^\dagger_{\mathbf{k};\sigma} & f^\dagger_{\mathbf{k};1\sigma}
    \end{pmatrix}
h^M_e(\mathbf{k})
    \begin{pmatrix}
        c_{\mathbf{k};\sigma} \\
        f_{\mathbf{k};1\sigma}
    \end{pmatrix}\\
=&\sum_{\mathbf{k},\sigma}
    \begin{pmatrix}
    c^\dagger_{\mathbf{k};\sigma} & f^\dagger_{\mathbf{k};1\sigma}
    \end{pmatrix}
    \begin{pmatrix}
        -h_c(\mathbf{k}) & \Phi \\
        \Phi & h_{f_1}(\mathbf{k}) 
    \end{pmatrix}
    \begin{pmatrix}
        c_{\mathbf{k};\sigma} \\
        f_{\mathbf{k};1\sigma}
    \end{pmatrix}\\
    =&\sum_{\mathbf{k},\sigma}
    \begin{pmatrix}
    c^\dagger_{\mathbf{k};\sigma} & f^\dagger_{\mathbf{k};1\sigma}
    \end{pmatrix}
    \left(
    \frac{-h_c(\mathbf{k})+h_{f_1}(\mathbf{k})}{2}+
   e^{-\mathrm{i}\tilde S_\mathbf{k}}
     \begin{pmatrix}
        0 & \Phi^\prime(\mathbf{k}) \\
        \Phi^\prime(\mathbf{k}) & 0 
    \end{pmatrix}
    e^{\mathrm{i}\tilde S_\mathbf{k}}
    \right)
    \begin{pmatrix}
        c_{\mathbf{k};\sigma} \\
        f_{\mathbf{k};1\sigma}
    \end{pmatrix},
\end{split}
\end{equation}
where $h_M^e(\mathbf{k})$ is the fourier transform of Eq.\ref{eqHMe} in the main text and,
\begin{equation}
\begin{split}
\Phi^\prime(\mathbf{k})=&\sqrt{\Phi^2+\left(\frac{h_c(\mathbf{k})+h_{f_1}(\mathbf{k})}{2}\right)^2},\\
    \tilde S_\mathbf{k}=&\frac{1}{2}\arctan\left(\frac{h_c(\mathbf{k})+h_{f_1}(\mathbf{k})}{2\Phi}\right)
    \begin{pmatrix}
        0 & -\mathrm{i} \\
        \mathrm{i} & 0
    \end{pmatrix}.
\end{split}
\end{equation}
We can define:
\begin{equation}
    \tilde S = \sum_{\mathbf{k},\sigma} 
    \begin{pmatrix}
    c^\dagger_{\mathbf{k};\sigma} & f^\dagger_{\mathbf{k};1\sigma}
    \end{pmatrix}
\tilde S_\mathbf{k}
    \begin{pmatrix}
        c_{\mathbf{k};\sigma} \\
        f_{\mathbf{k};1\sigma}
    \end{pmatrix}.
\end{equation}
Then for a given value of $\Phi$, we have $H^M_e=e^{-\mathrm{i}\tilde S}H^M_e(\Phi\rightarrow+\infty)e^{\mathrm{i}\tilde S}+\sum_{\mathbf{k},\sigma}\frac{-h_c(\mathbf{k})+h_{f_1}(\mathbf{k})}{2}(c^\dagger_{\mathbf{k};\sigma}c_{\mathbf{k};\sigma}+f^\dagger_{\mathbf{k};1\sigma}f_{\mathbf{k};1\sigma})$. The second term will not affect the ground state of $H^M_e$ as long as $\Phi$ is large enough to ensure that \textbf{there is exactly one eigenvalue of $h^M_e(\mathbf{k})$ less than $0$ for each $\mathbf{k}$}. Therefore,
the Slater determinant of $c$ and $f_1$ decided by $H^M_e$ can be obtained by performing a unitary transformation on Eq.\ref{eqGS1}, which can be written as:
\begin{equation}
    \begin{split}
        \ket{\mathrm{Slater}[c,f_1]}= & e^{-\mathrm{i}\tilde{S}}\prod_{i=1}^N\left(\frac{1}{2}(c^\dagger_{i;\uparrow}-f^\dagger_{i;1\uparrow})(c^\dagger_{i;\downarrow}-f^\dagger_{i;1\downarrow})\right)\ket{0}\\
        \approx & (1-\mathrm{i}\tilde S)\prod_{i=1}^N\left(\frac{1}{2}(c^\dagger_{i;\uparrow}-f^\dagger_{i;1\uparrow})(c^\dagger_{i;\downarrow}-f^\dagger_{i;1\downarrow})\right)\ket{0}.
    \end{split}
\end{equation}
Since we have:
\begin{equation}
\begin{split}
    1-\mathrm{i}\tilde S=&1-\frac{1}{2}\sum_{\mathbf{k},\sigma}\arctan\left(\frac{h_c(\mathbf{k})+h_{f_1}(\mathbf{k})}{2\Phi}\right)(c^\dagger_{\mathbf{k};\sigma}f_{\mathbf{k};1\sigma}-f^\dagger_{\mathbf{k};1\sigma}c_{\mathbf{k};\sigma})\\
   \approx &1-\sum_{\mathbf{k},\sigma}\frac{h_c(\mathbf{k})+h_{f_1}(\mathbf{k})}{4\Phi}(c^\dagger_{\mathbf{k};\sigma}f_{\mathbf{k};1\sigma}-f^\dagger_{\mathbf{k};1\sigma}c_{\mathbf{k};\sigma})\\
    =&1-\sum_{i,j,\sigma}\frac{t_{c,ij}+t_{1,ij}}{4\Phi}(c^\dagger_{i;\sigma}f_{j;1\sigma}-f^\dagger_{i;1\sigma}c_{j;\sigma}),
\end{split}
\end{equation}
 Therefore, the ancilla wavefunction in this case doesn't depend on $t_{c,ij}$ and $t_{1,ij}$ independently, but rather the sum of them. We can assume that $t_{c,ij}=t_{ij}$ and $t_{1,ij}=0$ for brevity. We use the relation Eq.\ref{eqcdf}, the final projected state can be calculated as:
\begin{equation}\label{eqGSphifinite}
    \ket{\Psi_c}=\left(1+\sum_{i,j,\sigma}\frac{t_{ij}}{2\Phi}c^\dagger_{i;\sigma}c_{j;\sigma}\right)\sum_{\{\sigma_i\}}\psi(\{\sigma_i\})\prod_{i=1}^N c^\dagger_{i;\sigma_i}\ket{0},
\end{equation}
which is the same as Eq.\ref{eqGSUfinite} if $\Phi=\frac{U}{2}$.
\begin{figure}[ht]
\centering
\includegraphics[width=0.5\linewidth]{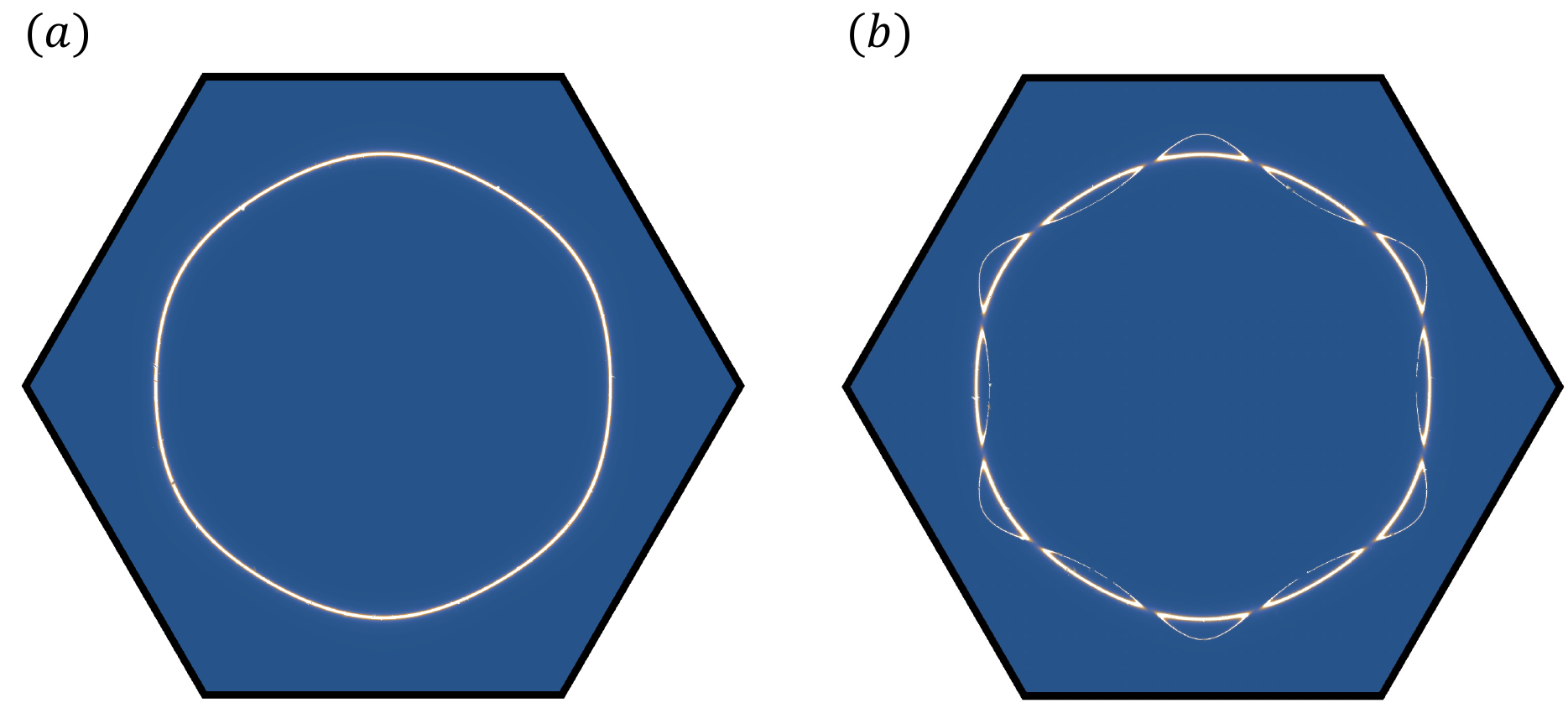}
\caption{The color plots of layer $f_1$'s spectral weight $A_{f_1}(\mathbf{k},\omega=0)$. (a) $\Phi=0$. (b) $\Phi=0.1$. }
\label{fig:spectralf1}
\end{figure}
\section{Calculation of the spectral weight function}
In this section, we show the details of the the computational details underlying the spectral weight presented in the main text. The spectral weight function of $c$ is defined as\cite{coleman2015introduction}:
\begin{equation}
    A_c(\mathbf{k},\omega)=\frac{1}{\pi}\mathrm{Im}G_c(\mathbf{k},\omega-\mathrm{i}\delta),
\end{equation}
where $G_c(\mathbf{k},\omega)$ is the Green's function of $c$ and $\delta\rightarrow0^+$. The Green's function in the free fermion system can be written as\cite{coleman2015introduction}:
\begin{equation}
    G_c(\mathbf{k},\omega)=\sum_{\lambda}\frac{|M_{c,\lambda}(\mathbf{k})|^2}{\omega-\epsilon_\lambda(\mathbf{k})+\mathrm{i}\delta\mathrm{sgn}(\epsilon_{\lambda})},
\end{equation}
where $\epsilon_\lambda(\mathbf{k})$ is the eigenvalue of the mean field Hamiltonian $h_M^e(\mathbf{k})$, $\lambda=1,2$. The corresponding eigenvector is $(M_{c,\lambda}(\mathbf{k}),M_{f_1,\lambda}(\mathbf{k}))^T$ and it is normalized to $1$. The spectral weight of $f_1$ is defined in a similar way.

In our calculation, we choose a small value for $\delta=0.005$, the spectral weight of $c$ is presented in Fig.\ref{Fig5} in the main text. The spectral weight of $f_1$ for the same hopping ansatz is presented in Fig.\ref{fig:spectralf1}.
\end{document}